\documentstyle[12pt,aaspp4]{article}

\input{epsf}
\begin{document}

\title{\bf X-ray outburst of CI Cam/XTE J0421+560: RXTE observations}

\author{Revnivtsev M.$^{1}$, Emelyanov A.$^{1}$, Borozdin K.$^{1,2}$}

\affil{$^{1}$ -- Space Research Institute, Moscow, Russia\\
$^{2}$ -- Los Alamos National Laboratory, Los Alamos, NM 87545, USA}

\begin{abstract}
We present the results of observations of XTE J0421+560 X-ray transient 
with Rossi X-ray Timing Explorer in the beginning of April 1998.
Lightcurve, obtained by ASM all-sky monitor, shows
unusually fast decrease in the object's luminosity. Spectra from series 
of observations by PCA and HEXTE experiments were studied in detail.
Two emission line regions with energies around 6.5 keV and around
8 keV have been mentioned in the spectra. The possibility of generation
of the observed emission in relativistic plasma jet has been discussed.
Some analogy is found with known Galactic jet source SS433.

\end{abstract}

\section{Introduction}

The X-ray transient XTE J0421+560 was discovered on Mar. 31, 1998 with 
All Sky Monitor aboard RXTE. According to ASM data the flux from 
the source in 1-12 keV energy range was rising quickly from 40 mCrab 
Mar. 31.36 to 1880 mCrab Apr. 1.04(Smith\&Remillard 1998). 
PCA and HEXTE observations started Apr. 1.08, 1998 and confirmed
the ASM results (Marshall\&Strohmayer, 1998). The new X-ray source was
localized by PCA with coordinates: Ra=$4^h19.6^m$, Dec=$+56^o00^{`}$
(Marshall\&Strohmayer,1998b). The lightcurve of the source was
characterized by extremely fast rise up to the maximum with subsequent,
also unusually fast decay of the flux. The observations by ASCA and 
BeppoSAX satellites showed the presence of the strong emission line 
around 6.7 keV.

The radio observations of Apr. 1, 1998 revealed the bright radio source at
the location   R.A.= $4^h19^m42^s.05$ +/- $0^s.03$, Dec. =
$+55^o59^{`}58^{``}.6$ +/- $0^{``}.5$, 2000 equinox (Hjellming\&Mioduszewski, 
1998a). The radio source, which position is
coincident with that of symbiotic star CI Cam (Wagner\&Starrfield, 1998) was
proposed to be the radio counterpart of X-ray transient XTE J0421+560. The
radio images obtained on Apr 5 and 6, 1998 revealed that the CI Cam radio
source had become resolved. The VLA images showed the extended radio emission
that had the appearance of a roughly symmetrical S-shaped twin-jet,
strikingly similar to the radio jets of SS 433 (Hjellming\&Mioduszewski,
1998c). The outermost pair of emission components was separated by
$\sim0.33^{``}$. Assuming the expansion started on Mar. 31.6, 1998 the
proper motion of the components is 26 mas/day. Assuming a distance of 1 kpc
(Chhikvadze 1970), this corresponds to apparent tangential velocity
$v\sim0.15c$.

Optical spectroscopy of CI Cam showed the strong emission lines of H, He I,
Fe II. However, none of the lines showed double peak profile, which is
typical for the accretion disk emission in X-ray binaries and cataclysmic 
variables. These results are consistent with
CI Cam/XTE J0421+560, but suggest that it might be unusual X-ray symbiotic 
star(Garsia et al., 1998).

\section{Observations and analysis}

In the present work we used the data obtained by Rossi X-ray Timing Explorer
(Bradt, Swank\&Rothschild, 1993). The data was retrieved from the XTE GOF at
GSFC. The ASM light curves were used as provided by the XTE GOF. The RXTE
has a payload of two coaligned spectrometers -- PCA and HEXTE -- that 
provides the broad band spectral coverage for energy range from 
3 to $\sim 200$ keV and All Sky Monitor, that give a possibility to follow
the long term behavior of the  source with nearly complete time coverage
within 1-12 keV energy band.

The PCA and HEXTE data was treated with standard FTOOLS v.4.1.1 package
tasks. To estimate the PCA background we run the latest version of task 
{\em pcabackest} v.2.0 (Stark 1998). We applied new $L7/240$ model to 
the observations, when the source was extremely weak (Apr. 8--9, 1998), 
and VLE based background model -- for others. In our spectral analysis 
we used PCA rmf matrix v.3.3 (Jahoda 1998, see also the $erratum$ 
Jahoda 1998a). The analysis of the Crab nebula spectra had confirmed, 
that the systematic uncertainties of the matrix are less than 1\%. 
We would like to note, that PCA effective area is substantially decreased 
after $\sim$3--4 keV to the lower energies, so the
estimation of $N_H$ value with PCA should be treated with care. 
To account roughly for the uncertainty in the knowledge of the
spectral response, 1\% of systematic error had been added to 
the statistical errors for each PCA channel. All spectra had been corrected 
for deadtime according to Zhang\&Jahoda, 1996.

For the spectral analysis of the HEXTE data the version 2.6 (released Mar.
20, 1997) of the response matrix had been taken. The background for each
cluster of HEXTE detectors was estimated using the off-source observations. 
Only the data above 15 keV were taken into account because of 
the uncertainties of the response matrix below this energy. 
At the high energy end the
spectrum was cut at $\sim 80-150$ keV depending on the brightness of
the source in order to avoid significant influence from the background 
subtraction uncertainty. The deadtime correction had been performed 
using hxtdead FTOOLS task for all observations. 

The brief information about the RXTE observations of the source are
presented in Table \ref{obslog}.

\section{Lightcurve}

\begin{figure*}
\epsfxsize=10cm
\epsffile{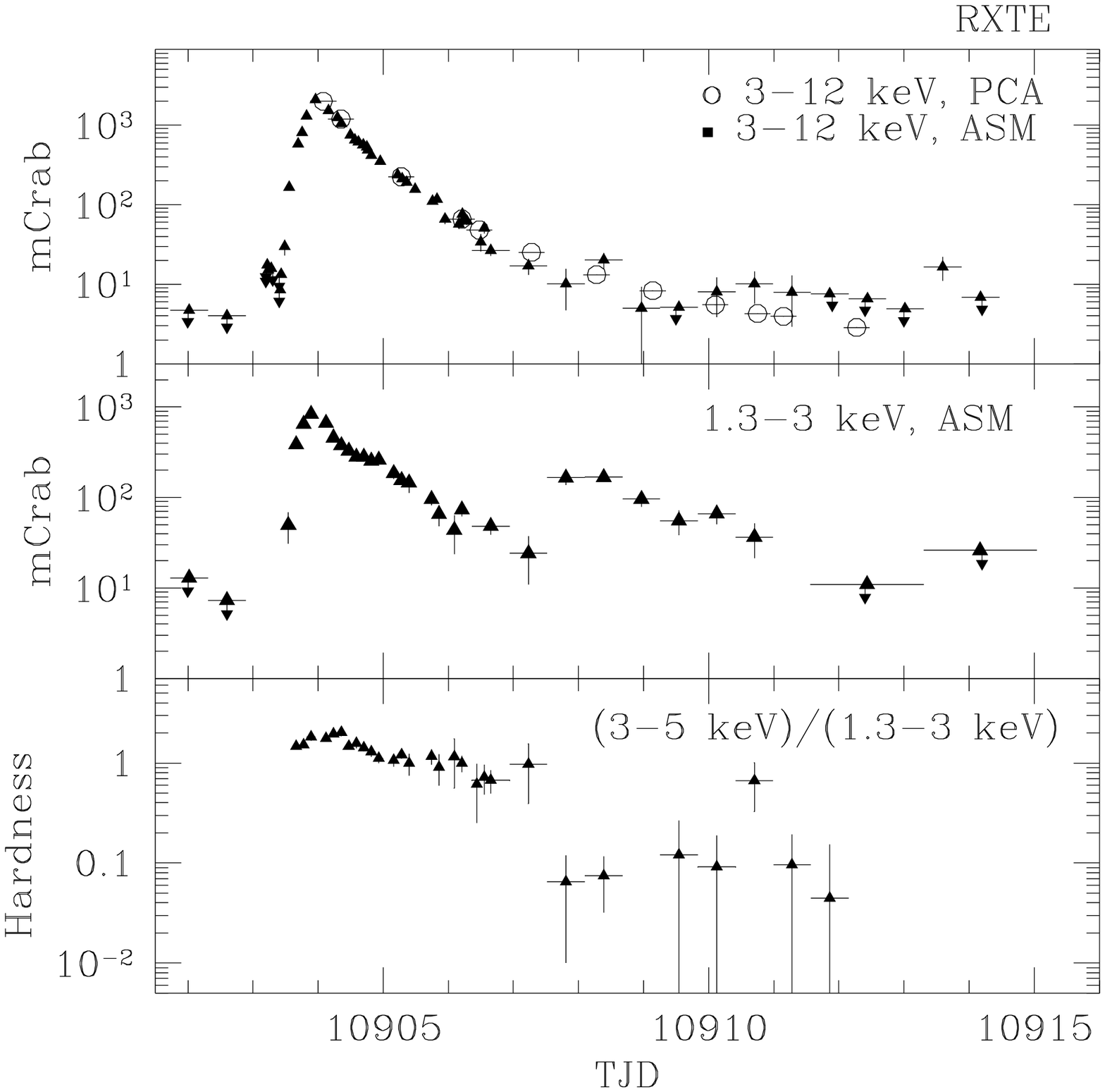}
\epsfxsize=10cm
\epsffile{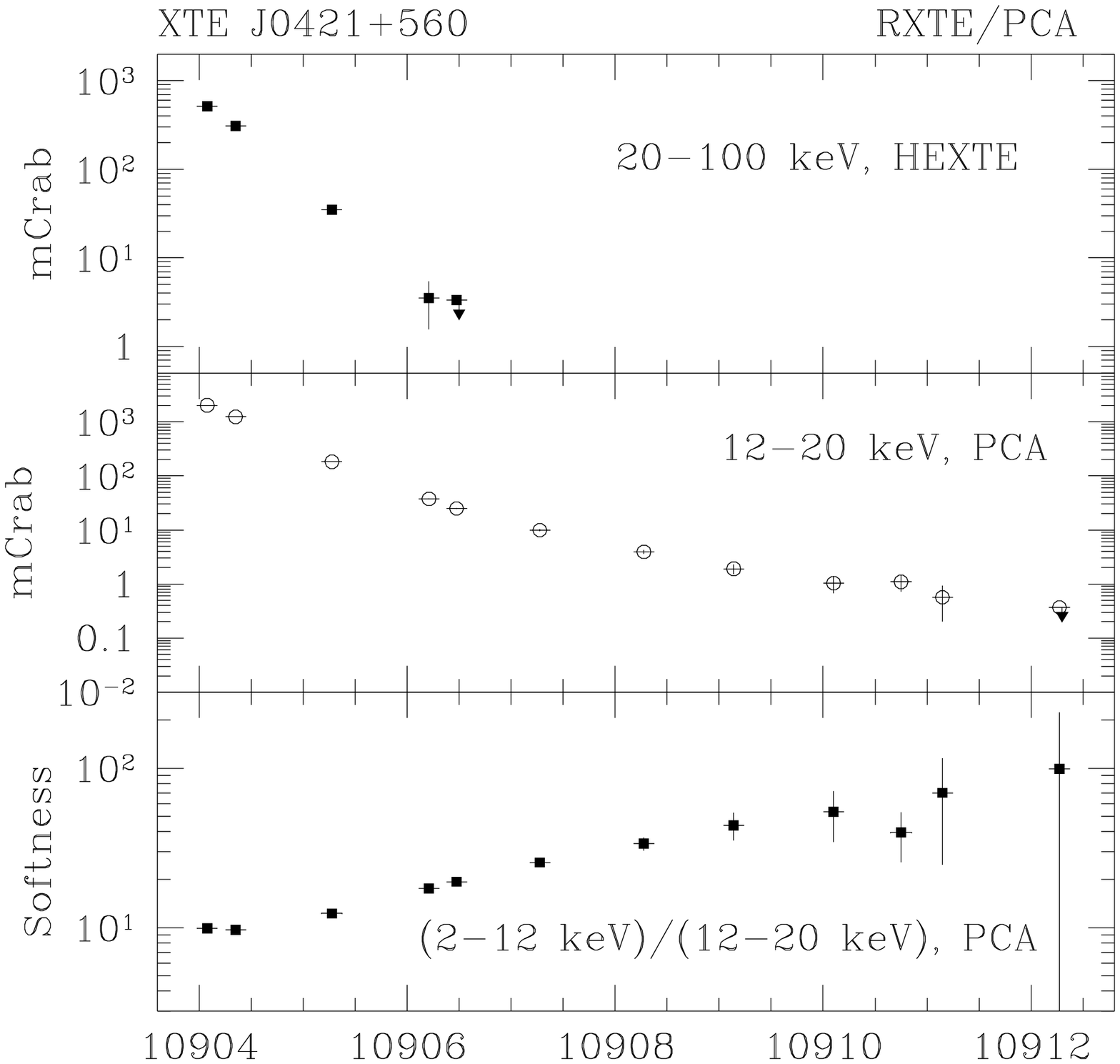}
\caption{The lightcurve of XTE J0421+560 according to RXTE data \label{lcurve}}
\end{figure*}

The lightcurve of the source in various energy bands is shown in Fig.
\ref{lcurve}. The peculiar point in this figure is that the ASM data in the
lowest energy band 1.3-3.0 keV show the feature that looks like secondary
peak around the 5th day after the beginning of the outburst
(Fig.\ref{lcurve}a). Similar ``kick'' feature is frequent for the lightcurves
of the X-ray novae in standard X-ray band, but 
in our case the feature is seen only in the soft X-ray band ($<$ 3 keV)
while the flux at higher energies is decreasing monotonically . Note, that
flaring in softer energy band was observed also by ASCA observatory around 
the same date (Ueda et al., 1998b)

\begin{figure*}
\epsfxsize=16cm
\epsffile{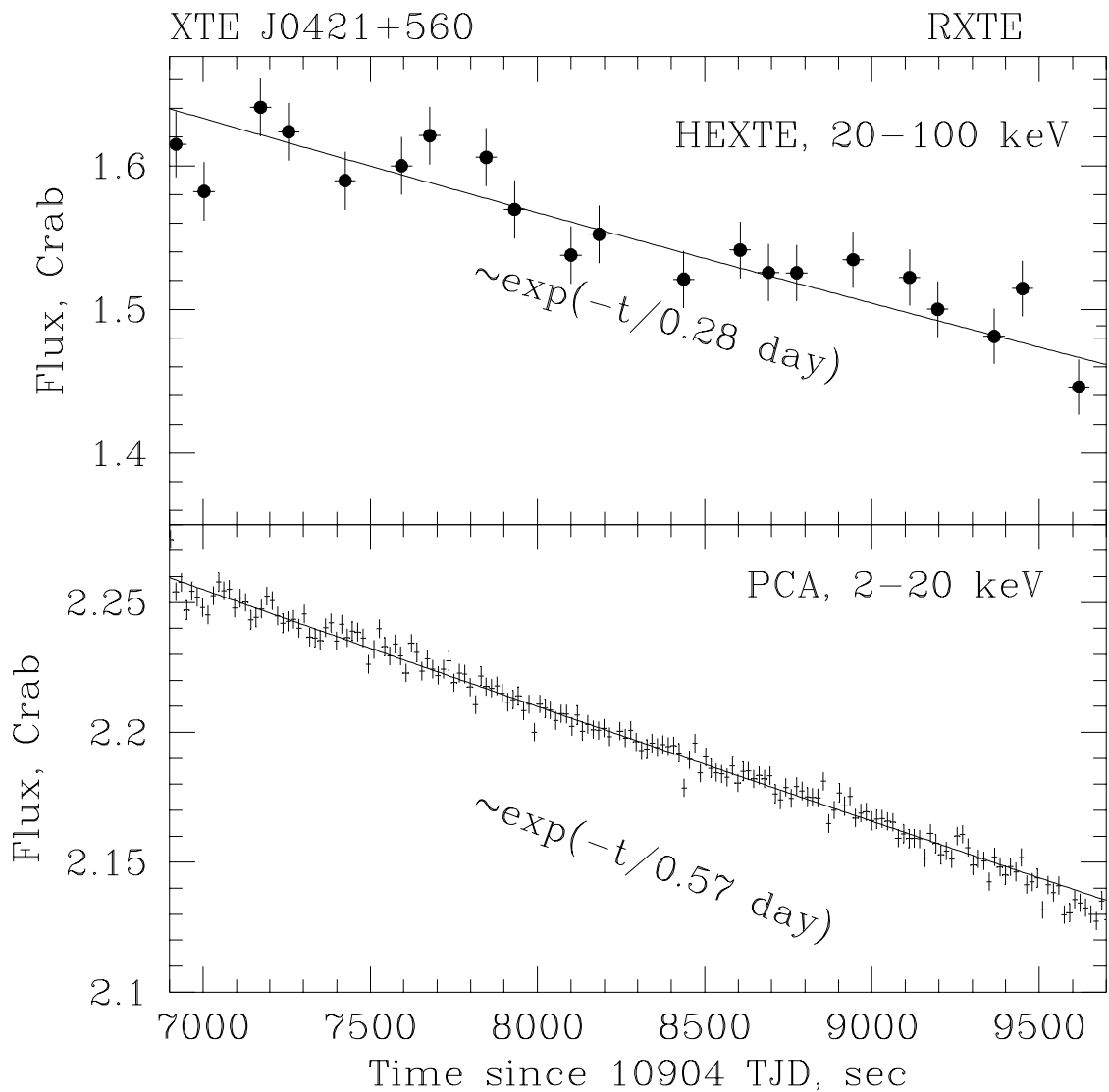}
\caption{The lightcurve of XTE J0421+560 during the part of the first
observation (PCA and HEXTE data). The solid lines show approximation
$e^{-{{t-t_o}\over{\tau}}}$, where $\tau\sim$0.56 day for PCA points (lower
panel) and $\tau\sim$0.28 day for HEXTE points (upper panel), $t_o$ was
taken to be Mar. 31.6, 1998. One can see
that the flux from the source do not demostrate significant variability on
the time scales 20--1000 s. The $\chi^2$ value for the PCA points -- 182 for
162 points.\label{pca_hexte}} 
\end{figure*}

In the lower panels the hardness ratios of the source are presented. The
flux decay during the period Apr.1-3 in the total ASM energy band can be
approximated with the exponential law with e-fold time  $\sim$0.6 day, 
later on the e-fold time increases to $\sim$1.1 day.  The example 
of the source flux decay during the first session of observation 
by the PCA and HEXTE is shown in the Fig.\ref{pca_hexte}.
During this observation e-fold times were $\sim$0.57 day for the PCA
energy band and $\sim$0.28 day for the HEXTE energy band.
 
\section{Energy spectrum}

\begin{figure*}
\epsfxsize=16cm
\epsffile{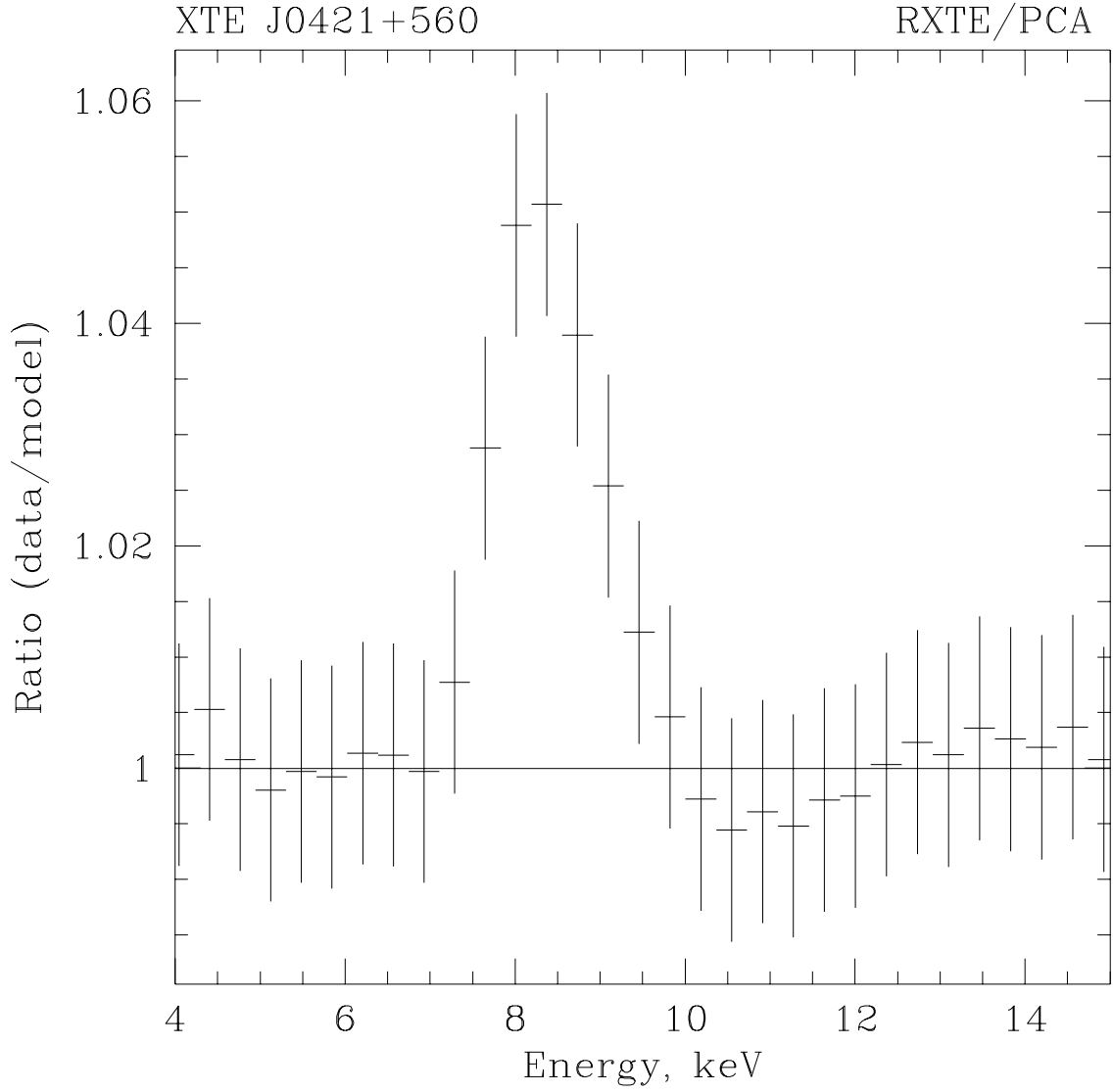}
\caption{The ratio data/model, when 8 keV line was not taken into account.
Systematic uncertainties of 1\% are shown.\label{8keV}}
\end{figure*}

The source spectrum in the 3--100 keV energy band can by approximated by a
power law with the exponential cutoff at higher energies: $dN \sim
E^{-\alpha} e^{-{E\over{E_{cut}}}}dE$. Extremely strong emission line 
is present in the spectrum at energies around 6.4--6.7 keV. 
Similar spectral shape was found by ASCA and BeppoSAX observatories 
for their observations (Ueda et al., 1998, Orlandini et al., 1998). However,
for the PCA spectra this model leaves the significant residual at 
the energies around 8 keV with the amplitude ($\sim$4--6\%, see
Fig.\ref{8keV}), which is  
sufficiently higher than the response matrix uncertainty. Because of 
the presence of very bright emission line at the energies 6.4--6.7 keV 
and low energy resolution of PCA in this energy band ($\sim$1 keV), 
the quality of the data approximation at the energies $\sim$8
keV will strongly depend on the accuracy of used photon redistribution
matrix (definite part of the response matrix). The analysis, carried out
with collaboration of Keith Jahoda (RXTE/PCA team) have showed, 
that the unaccuracy of the matrix is not worse than $\sim$1\% and 
the observed excess around 8 keV should be considered as real 
intrinsic spectral feature of the source. An addition of the emission 
line at energy $\sim$8.1 keV significantly improves the quality of
the data approximation in terms of $\chi^2$ statistics -- $\chi^2$ value
reduces by $\sim$40 with the systematic uncertainties included and by
$\sim$9000 with statistical errors only. As a result, the best fit analytical model 
consists of a power law with high energy exponential cutoff, low energy 
absorption, and two emission lines. The parameters of this model 
for different sessions are given in Tables \ref{cutoff} and \ref{lines}. 
We would like to note, that the spectral results of PCA with response 
matrix 3.3 have relative uncertainties in energy around 0.02 keV 
at the energies 6--8 keV, but the uncertainties in absolute values 
is slightly higher, it can be estimated
to be $\sim1\%\sim0.07$ keV (Jahoda 1998, private communication). One 
should bear this in mind, while tring to localize the lines in the
spectrum.

Combined spectra of PCA and HEXTE experiments for three different
observations are presented in Fig.4 (in the Fig.4 $meka$ model was used to
approximate the line emission, see below). The spectra can be approximated
satisfactorily by the aforementioned model. Note, hovewer, 
that because of calibration 
uncertainties the photon indexes of the Crab spectra of PCA and HEXTE 
have slightly different values (on $\sim$ 0.05--0.1), 
therefore some deviations in spectral shapes between two experiments 
can be expected.

Due to the decrease of the flux and the softening of the spectrum 
from XTE J0421+560 the source had not been detectable by HEXTE 
after Apr.4, 1998.

The shape of the spectra is close to the spectrum of the emission
of optically thin plasma ($bremsstrahlung$), but cannot be satisfactorily
fitted by this model with single temperature.  Possible 
interpretations of this fact could be fast change of the temperature
affecting the intergral spectrum or non-equilibrium state of the
emitting plasma.
 
It is worth to note several features of the obtained spectra:
1) the absorbtion value $N_H$, that characterizes the low energy cutoff in
the spectra is changing from $\sim4.8\times10^{22}$ on Apr.1.04,
$\sim6.5\times10^{22}$ Apr.1.35, to $\sim1.5\times10^{22}$ the day after and
then became insignificant;
2) the slope of the power law fit increases and the cutoff energy decreases 
with time, which shows the cooling of the emission medium;
3) the center of the broad emission line 6.5--6.7 keV fluctuates to 
the higher energies. (see Fig. \ref{6keV} and \ref{lines}).

Below we discuss the possible interpretation of these and other
observational results.

\section{Discussion}

The X-ray transient source XTE J0421+560/CI Cam has several featues that
make it different from the other known X-ray transients. The most striking
difference, which has been repeatedly mentioned in the related publications,
is an extremely fast evolution. Mean value of e-folding decay time for
know X-ray transient is $\sim$ 30 days (see Wan Chen et al. 1997),
whereas for XTE J0421+560 it is $\sim$ 0.5--1 day. Possibly, the main
difference is the nature of the compact object in these systems. CI Cam is
known to be symbiotic binary, i.e. the binary which composed by red giant
and an accreting white dwarf.  The most probable cause of
the flares in these systems is considered to be the thermonuclear 
outburst on the surface of the white dwarf. Unstationary
accretion in the system is proposed as an alternative mehanism for
the flares generation (see e.g, review of the models by 
Mikolajewska\&Kenyon 1992).

It should be noted that the X-ray observations of XTE J0421+560 do not
demonstrate the features, which is usually attributed to the presence of the
accretion disk in the system, -- QPO, or characteristic Very Low Frequency
Noise in the power spectrum, or typical soft component in the energy
spectrum. Moreover, the analysis of the light curve of the source on the
various time scales (from hundreds of microseconds up to hundreds and
thousands of seconds) do not reveal the significant variability of the
source flux different from the main trend of the flux decay. This result is
consistent with observations of ASCA and SAX (Ueda et al., 1998b, 
Frontera et al.,1998). Hovewer, according to SAX and ASCA data X-ray 
flux in the 0.5-1 keV energy band do has variability on the scales 
of hundreds of seconds and hours.

The character of the source flux variability could be an evidence that 
the soft and hard X-ray components are forming in the geometrically 
different parts of the system. The dimension of the region, where 
the hard component is formed, should be larger than 
$R\sim c_s\tau\sim10^8\times100=10^{10}$ cm, here
$c_s$ -- the sound velocity in the medium with $T\sim10$ keV, while 
the region, where the soft component is formed, must be less than 
$\sim10^{10}$ cm in size.

\subsection{Continuum spectrum}
As it has been mentioned above, the best fit model for continuum spectrum 
is the power law with high energy cutoff at the energies $\sim$5--13 keV. The
spectrum of this type can be formed in the cloud of optically thin hot
plasma, that has non-uniform temperature distribution. It should be noted
that this form of the spectra is completely different from that of the many
other X-ray Novae - black hole candidates, where X-ray emission was detected
up to hundreds keV (see e.g., Sunyaev et al., 1994; Tanaka\&Shibazaki, 1997).

\subsection{Low energy absorption}

The spectra, obtained during the first two days after the beginning of the
outburst demonstrate the presense of significant low energy absorption.
The $N_H$ value was increased from $N_{HL}\sim4-5\times 10^{22}$ on Apr.1.04, 
1998 (model dependent value) to $N_{HL}\sim6-7\times 10^{22}$ on Apr. 1.35, 
1998 and then decreased to $N_{HL}\sim1-2\times 10^{22}$ one day later
and had become insignificant during later observations. It is quite 
natural to interpret that at the beginning of the outburst the X-ray
source had been embedded within dense and cold cloud, which later on became
less dense and, consequently, more transparent as a result of mass loss 
and/or temperature heating. Lightcirves of the object, obtained in X-ray, 
optical and radio bands may also be considered as an evidence of the cloud
density decrease with time(Frontera et al.,1998).

The radio observations of the relativistic jet structures 
(Hjellming\&Mioduszewski, 1998c) supports an idea of significant outflow 
of the matter from the central object of the system XTE J0421+560
during the outburst.  It is interesting to estimate the dimensions 
and the density of the cloud assuming its cylindrical shape.

Let us assume, that the cloud forms a cylinder with radius $R$, length $L$
expanding along the cylinder axis, perpendicular to the line of sight 
(see below the discussion of orientation of the jets). 
We accept for simplicity that the cloud has the uniform
distribution of the density. The low energy absorption observed can be
caused either by bremsstrahlung self absorption or by photoabsorption in
the cool outer region of the cloud. With soft X-ray instrument in hand it
would be possible to distinguish between these two possibilities because of
its different influence on the source spectrum. Unfortunately, PCA do not
allows to do it.
 
If the low energy cutoff in the 
spetra of XTE J0421+560 is due to photoabsorption, then we can make 
some simple estimations on the cloud dimensions.
The value of photoabsorption can be approximated as $N_H\sim C NR$, 
where $C<1$-- part of the cloud which contribute to the $N_H$ value, 
$N$-- density of the cloud. Emission measure in the cloud $EM$ is given by:
\begin{equation}
EM=\int{N^2dV}=N^2V \sim N^2 \pi R^2L.
\end{equation}

Then we obtain:
\begin{equation}
L\sim{C^2 EM\over{N_H^2}} \sim {C^26\times10^{59}\over{(5\times10^{22})^2}} 
\sim 2.4C^2\times10^{14} cm.
\end{equation}

Taking into account that this value is just order-of-magnitude evaluation 
of cloud length we can say, that our estimation is consistent 
with the observed expansion rate of the cloud in this direction  
$\sim1\times10^{13}$ cm/hour if $C\sim0.1-0.2$. It should be noted, 
that if the radiation goes through the
cold surrounding medium, the weak fluorescent line of Fe at the energy
$\sim6.4$ keV should be formed (see the discussion of this item below 
and also in Ueda et al., 1998b). We emphasize that the discussed model 
should not be considered as ultimate, but as one of the possible ones.

Under assumptions that there is no energy supply to the cloud after the maximum
of the lightcurve is reached, and that to the moment of sufficient 
flux decrease (after $\sim$3--4 days after the peak) the cloud have lost 
a significant part of its energy, we can estimate the density of the cloud. 
Let us take the characteristic time of the flare to be $\tau\sim$1--2 days 
($\sim10^5$ s). From the energy balance of the cloud 
(see also Ueda et al., 1998b):
\begin{equation}
{NkT\over{\tau}}\ga A N^2 \sqrt{T}, A\sim1.7\times10^{-27} erg/s
\end{equation}
and
\begin{equation}
N\la{k\sqrt{T}\over{\tau A}}\sim10^{10} cm^{-3}
\end{equation}

Accepting this value of the density, we can evaluate the transverse size of the
cloud $R \sim N_H/NC \sim 0.5-1.0\times 10^{13}$ cm and total mass of the
erupted matter $M\sim m_pNV \sim m_pEM/N \sim 8 \times 10^{25}$g$ \sim
4\times10^{-8} M_{\odot}$. The obtained estimated value of $R$ is consistent
with the free expansion of the cloud in this direction $R\sim c_s \tau \sim
10^{13}$ ($c_s$ -- the sound velocity of the gas with the temperature
$kT\sim10$ keV).

\subsection{Emission lines origin}

The most likely, that the observed lines in the spectrum of XTE J0421+560
has been generated in the optically thin emission of the hot plasma cloud. 
Alternative mechanisms of line formation -- the propagation though 
the medium or the reflection -- is less probable, because of huge 
equivalent width of the observed line. Besides, if the propagation 
or the reflection would be responsible for the line formation then 
the strong absorption edge (EW$>700$ eV) would be formed at energies 
above the line. The lack of such a feature moves us to believe that 
both the continuum spectrum, that extends up to the 100 keV 
and the observed emission features are formed
in the hot optically thin plasma cloud. However, the PCA data show that the
model of emission of optically thin plasma with lines with $kT\sim$10--15
keV doesn't give a good approximation of the spectra ($\chi^2=187.8$ 
for 41 d.o.f for the first observation). High $\chi^2$ value is mainly 
driven by the fact that the model fails to
predict the right ratio between the fluxes detected from different lines. 
If we take two-component model that consists of the emission of optically thin
plasma cloud without emission lines ($bremsstrahlung$) with $kT\sim$14 keV
and the emission of the plasma cloud with emission lines ($meka$) with
$kT\sim$5 keV, then we will get much better fit in terms of
$\chi^2$ ($\sim16$ for 38 d.o.f.). It should be noted that
in order to get the good approximation one should take into account the
Doppler shift $z\sim 0.03$ (see. Table \ref{bremss_meka} and the discussion
below). So, we believe that the spectrum of XTE J0421+560 consists of (at
least) two components, one of which describes the continuum spectrum extended 
up to $\sim$100 keV and the second one represents the region, where
emission lines are formed. The difference between the characteristic
temperatures of these two components during whole outburst may hint on the
fact that they are emitted by the geometrically different regions of the
system (see also analysis of the flux variability above). The best fit
parameters for the discussed model are presented in Table \ref{bremss_meka}.
The combined PCA and HEXTE spectra shown in Fig.4. 

\subsection{The emission line positions in the spectra}
     
The PCA data demonstrate that the spectrum of the source has strong emission
line features at the energies around 6.5--6.7 keV and around 8 keV. 
The existence of the strong line at the energy $\sim$6.7 keV has been
reported by all instruments which observed XTE J0421+560 and had needed 
capabilities in this energy range. New and interesting results of
the present work is the shift of the central energy for Fe-line complex 
and the detection of additional significant line-shaped emission around 8
keV \footnote{As it was mentioned above, 8 keV line emission can be well
approximated by $meka$ model}.
In the light of recent radio observations of relativistic jets it looks 
attractive to interpret the observed drift of Fe-line in terms of the change
of Doppler effect value for relativistically moving plasma.  This
interpretation is in agreement with results obtained by ASCA.  Estimated
velocity of the expansion is equal to $\sim0.03c$ for the observation
on April 1, with following decrease of this value leading to observed
shift of line centrum to higher energies.  If an initial shift is really
caused by Doppler effect, then an orientation of the cloud must be equal
to $\sim arctg(0.15/0.03)\sim 80^o$ relative to the line of sight.
Thus the expected shape of the jets should be close to symmetrical as
it was really observed in radio band.  Our expansion velocity estimation
is in principal agreement with ASCA results (Ueda et al., 1998b).

One serious argument against this interpretation is the lack of 
blue-shifted line component in the spectra.  Emission line $\sim$ 8 keV
is hardly can be interpreted as such a component, because, even taking 
into account an uncertainty in its position, its intensity is still
much weaker than for 6.5-6.7 keV line.  Hovewer, one should take into 
account that PCA energy resolution ($\sim 18\%$) does not allow to
distinguish between multiple lines which might present in this energy
band.

It is remarkable, that similar shift in emission line energy up to
$\ga7$ keV had been observed for SS433 source during its observation 
by EXOSAT and had been interpreted in terms of Doppler effect (Watson et al., 
1986). But in that case only blue-shifted component had been detected,
and the absense of red-shifted line was explained for the screening 
of the emitting region by the accretion disk.

If the change in the energy observed for 6.7 keV emission line is due to
the relativistic motion of an emitting cloud of optically thin plasma,
then the decrease in $\Delta E$ with time can be attributed to the
decrease of the velosity of the cloud or to the precession of the jet axis,
in the analogy with SS433 system.  Hovewer, the time dependence of 
the emission measure (see below) does not show any significant decrease
in the expansion velosity, which is an argument in favor of the precession.
In this case a period of the precession is $>10$ days (in comparison
with $\sim164$ days for SS433). S-shaped structure observed in radio
band is another argument, which suggest that the change in observed
expansion velosity is due to the change of the projection of the expansion 
vector to the line of sight.

As an alternative explanation of the observed shift of the line to lower
energies one can consider a presence in the spectrum of fluorescent Fe-line
at 6.4 keV, which is generated during the passage of X-rays through
cold dense cloud surrounding the X-ray source. It is possible that
an equivalent width of this line has decreased significantly during
later observations, which resulted to the shift of the centrum of broad
emission feature detected by PCA.  The trend of low-frequency absorption
discussed above is in favor of this interpretation.  Hovewer, it looks 
like higher optical depth of cold cloud is needed to form 6.4 keV line
with sufficient width.  This model is also discussed by Ueda et al.(1998) 
as a possible interpretation of ASCA results.

The spectral feature observed around 8 keV could be attributed to the
emission line of highly ionized Ni or to Fe $K_{\beta}$ line complex.
While an emission in Fe line with energies $\sim6.7$ keV has been
widely observed from many X-ray sources, the detection of Ni-line is
quite unusual result. We believe, that this detection became possible
in this case only because of extreme brightness of the source and unusually
high significance of line emission component in its spectrum combined 
with high sensitivity of PCA experiment.

\subsection{Evolution of X-ray spectra}

To trace the changes of physical parameters for an emitting system we
have applied two-component model compoused of the component of optically
thin plasma emission without lines and another component of
optically thin plasma with line emission (as has been discussed above).
We have used PCA data only, because its sensitivity is high enough to
trace spectral parameters change for small exposure times.  We find 
best fit parameters for spectra integrated over 400-sec time intervals 
for the first three observational sessions,
when the brightness of the source have been at maximum, 
and one integrated spectrum for the rest of the sessions.
The dependencies obtained for emission measure and temperature parameters
are shown in Figures 7 and 8. It is evident from Fig.7 that after 
$\sim 30$ hours after the beginning of the outburst emission measure
for both components is decreasing as $EM \sim t^{-2}$, while overall
behaviour of this parameter can be described as 
$EM\sim (1+(t/t_0)^{\alpha})^{-1}$, where $\alpha\sim$ 2.17 for
the first component (bremsstrahlung without lines) and 
$\alpha\sim$ 2.25 for the second component with lines (meka).
There is no evidence in Fig.7 that expansion velosity is decreasing
with time. 

If we assume that cloud mass $M\sim m_pNV$ is constant, then its
internal thermal energy will depend of the temperature T only, and
emission measure - of density N only. In this case the volume of
the cloud is proportional to the emission measure: $V\sim EM^{-1}$.
From the observations the time dependence of the emission measure
parameter is given by $EM\sim (1+({t\over{t_0}})^{\sim 2})^{-1}$.
Then the cloud volume will increase with time as $V\sim (1+({t\over{t_0}})^2)$.
We note, that index value $\alpha\sim 2<3$ is an indication of
non-isotrop or collimated expansion of the cloud.

The temperature vs time dependence is quite complicated (Fig.8), 
but one can mention two main features to describe it.  During
the first 20-30 hours after March 31.6 the temperature of the
components remain relatively constant in comparison with significant
change of the emission measure. After $\sim 30$ hours, the temperature
is decreasing as $\sim t^{-0.6}$.

\subsection{The comparison with other X-ray sources}

While XTE J0421+560 have a set of prominent differencies from
other X-ray transients, radio observations have shown an evident
similarity between this source and SS433, well-known Galactic
object emitting relativistic plasma jets. An analogy between two
sources is revealed itself in X-ray band also, because both of 
them have similar continuum spectra and strong emission lines
aroun 6.7 keV. Furthermore, the shift of line centra with time is
detected for both sources which might be explained in terms of
Doppler shift in line energy. The important difference between
two sources is a fact that XTE J0421+560, contrary to SS433, is
a transient object, which had been active in X-ray and radio bands
for few days only.

We would like to note, that rapid decrease of X-ray flux with 
simultaneous softening of the spectrum, as well as shift of
the maximum of energy emission with time for different bands of
electromagnetic spectrum is typical also for gamma-burst afterglows.

\section{Conclusion}

The observations of X-ray transient XTE J0421+560, which is an X-ray
counterpart of symbiotic binary CI Cam, by RXTE experiments give
a set of important and unexpected results.  First of all, the source
attracted the attention because of unusually fast rise of its X-ray
flux, which was followed by also too fast decline after the maximum
passage. During 10 days of observations by PCA and HEXTE experiments
the flux from the source had been decreased by more than two orders
of magnitude.

One interesting result is the detection of soft X-ray flare, which
was not correlated with the flux change in harder energy range.

The source spectrum in 3-150 keV band can be approximated by
power-law with an exponential cut-off at higher energies.
The spectrum of this shape can be generated in the cloud of
non-thermalized optically thin plasma. The softening of the spectrum
(or decrease of the effective temperature) with the decrease of
flux has been mentioned. 

The changes in low-energy absorption in the spectrum is probably
due to the decrease in density of the cold cloud, which surrounds
X-ray source. Initial mass of the cloud can be estimated as 
$\sim 4 \times 10^{-8}M_{\odot}$.

The emission lines have been detected at energies 6.5-6.7 keV and
around 8 keV. Discovered shift of the 6.7-keV line to lower energies
can be explained either as Doppler-effect for moving media, or as
generation of another, 6.4-keV emission line in cold cloud, density
of which evolved with time.  If the reason is Doppler-effect,
then this is the second time, after SS433, when an X-ray emission 
generated by relativistic jet plasma is observed in our Galaxy.
In this case the cloud is expanded under angle $\sim80^o$ to the
line of sight.  The appearance of the line near 8 keV is probably
due to the emission of highly ionized Ni or Fe $K_{\beta}$ line,
and, as far as we know, is detected for the first time for Galactic
sources.

\medskip

Authors thank E.Churazov for helpfull advises and comments, S. Sazonov, S.
Trudolyubov and A. Finoguenov for extensive discussions of the presented
results. Authors especially thank Keith Jahoda for the help in the analysis
of spectral capabilities of PCA. Also we thank Y.Ueda and H.Inoue for the
possibility to get to know ASCA results before its publication.
The research has made use of data obtained through the High Energy
Astrophysics Science Archive Research Center Online Service, provided by the
NASA/Goddard Space Flight Center. This work was partly supported be RBRF
grant 96-15-96343.
 
\section{References}

\indent

Bradt H., Rotshild R., Swank J.
\small Astron. Astrophys. Suppl. Ser., 1993, vol. 97, no. 1, p. 355-360.

Chen Wan, Shrader C., Livio Mario
\small Astrophys. J., 1997, v.491, p.312.

Chkhikvadze Ja.
\small Astrofisika 6, 65, 1970.

Frontera F., Orlandini M., Amati L., Dal Fiume D., Masetti N., Orr A.,
Parmar A., Brocato E., Raimondo G., Piersimoni A., Tavani M., Remillard R. 
\small Astron. Astrophys. 1998, see also astro-ph/9809287

Hjellming R., Mioduszewski a.
\small IAU Circ. No. 6857, 1998a

Hjellming R., Mioduszewski b.
\small IAU Circ. No. 6862, 1998b

Hjellming R., Mioduszewski c.
\small IAU Circ. No. 6872, 1998c.

Jahoda K.
\small 1998a, http://lheawww.gsfc.nasa.gov/ users/ keith/ pcarmf.html

Jahoda K.
\small 1998b, http://lheawww.gsfc.nasa.gov/users\\/keith/pcarmf\_ft41.erratum

Marshall F., Strohmayer T.
\small IAU Circ. No. 6857, 1998

Mikolajewska Joanna, Kenyon S.
\small MNRAS 1992, vol. 256, no. 2, p. 177-185.

Orlandini M., Dal Fiume D., Frontera F., Antonelli L., Piro L., Parmar A.
\small IAU Circ. No. 6868, 1998

Smith D., Remillard R.
\small IAU Circ. No. 6855, 1998

Stark
\small 1998, http://lheawww.gsfc.nasa.gov/ $\sim$stark/ pca/ pcabackest.html

Sunyaev R., Borozdin K.,Aleksandrovich N., Arefev V., Kaniovskii A., Efremov
V., Maisack M., Reppin C., Skinner J.
\small Astr. Lett. vol.20, no.6, p. 777, 1994

Ueda Y., Ishida M., Inoue H., Dotani T., Lewin W.H.G., Greiner J.
\small  IAU Circ. No. 6872, 1998a.

Ueda Y., Ishida M., Inoue H., Dotani T., Greiner J., Lewin W.H.G.
\small Astrophys. J. 1998b, see also  astro-ph/9810100.

Wagner R., Starrfield S.
\small IAU Circ. No. 6857, 1998

Watson M., Stewart G.,Brinkmann W., King A.
\small MNRAS 1986, 222, 261-271.

Zhang W., Jahoda K.
\small 1996, http://lheawww.gsfc.nasa.gov/ users/ keith/ deadtime/deadtime.html

\large
\clearpage

\begin{figure*}
\epsfxsize=16cm
\epsffile{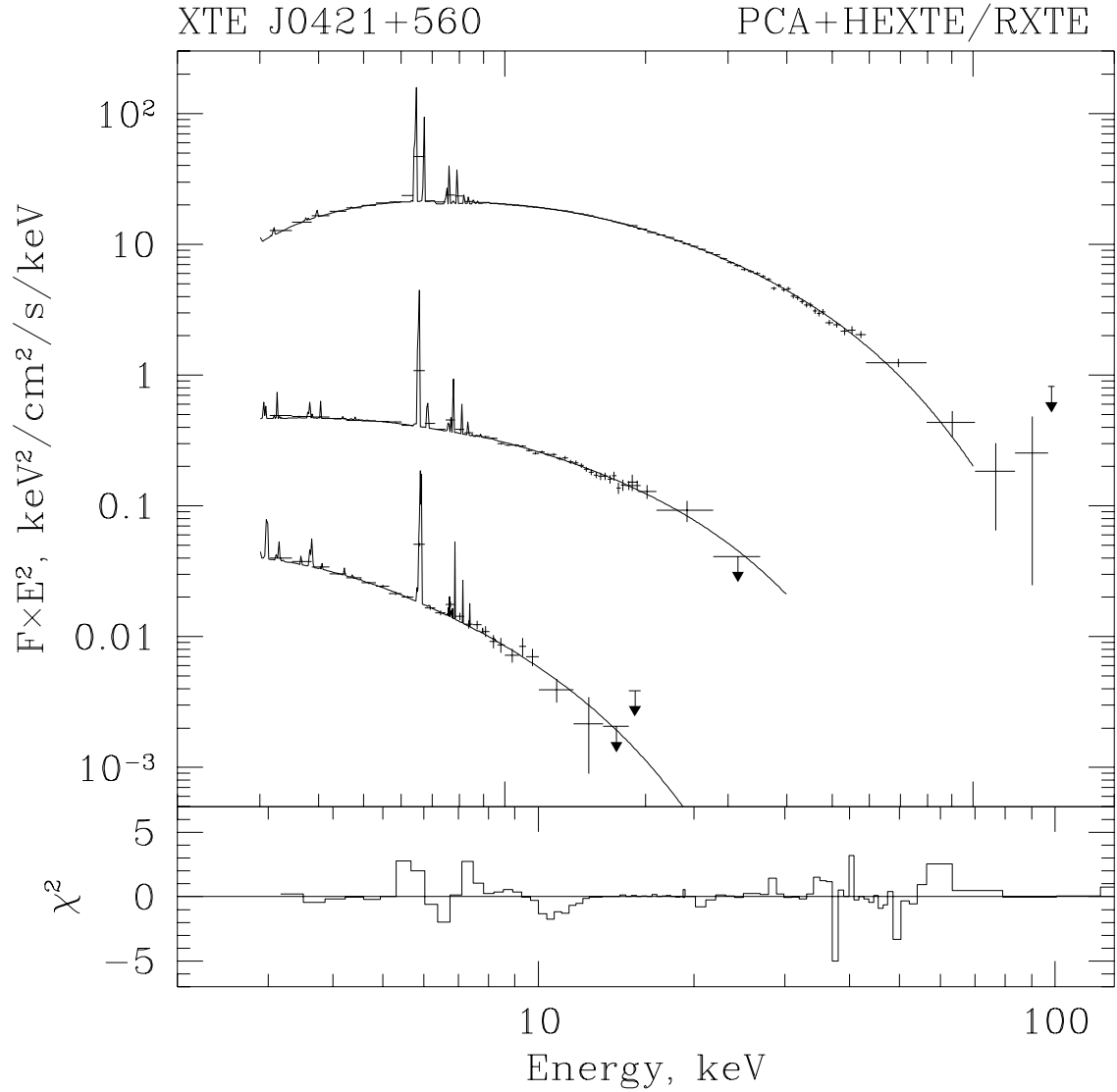}
\caption{Spectra of XTE J0421+560 in different luminosity states. The
most luminous one -- on Apr. 1, 1998, the middle one -- Apr. 3, 1998, the
weakest -- Apr. 9, 1998. On the lower panel the $\chi^2$ curve for the upper
spectrum is presented. Not all of shown lines are seen by PCA, a number of
emission lines caused by $meka$ model (see text). \label{spectra}}   
\end{figure*}

\begin{figure*}
\epsfxsize=16cm
\epsffile{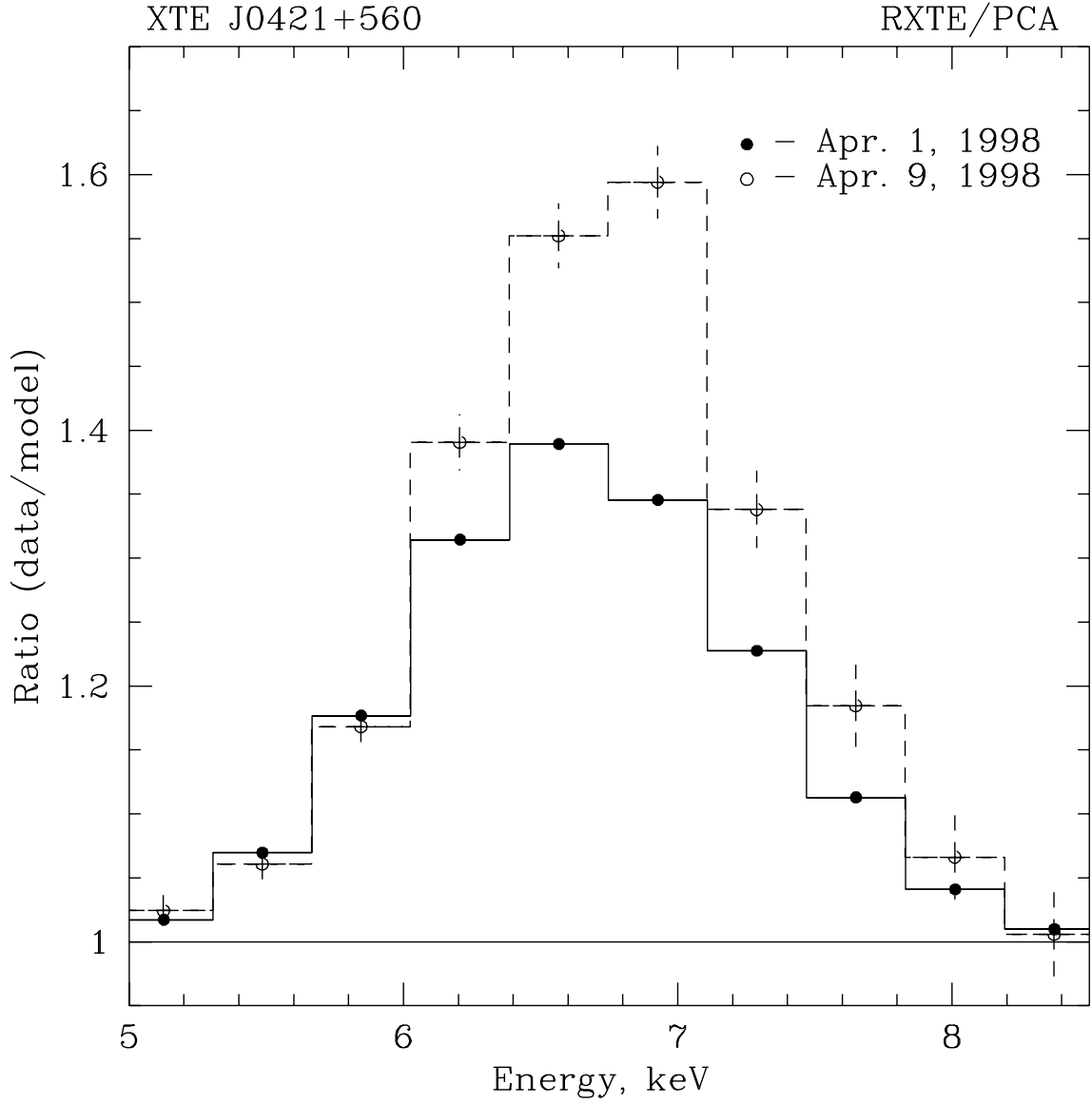}
\caption{The ratio data/model when 6.5--6.7 keV line was not taken into
account. One can see the shift of the line central energy.\label{6keV}}
\end{figure*}

\begin{figure*}
\epsfxsize=16cm
\epsffile{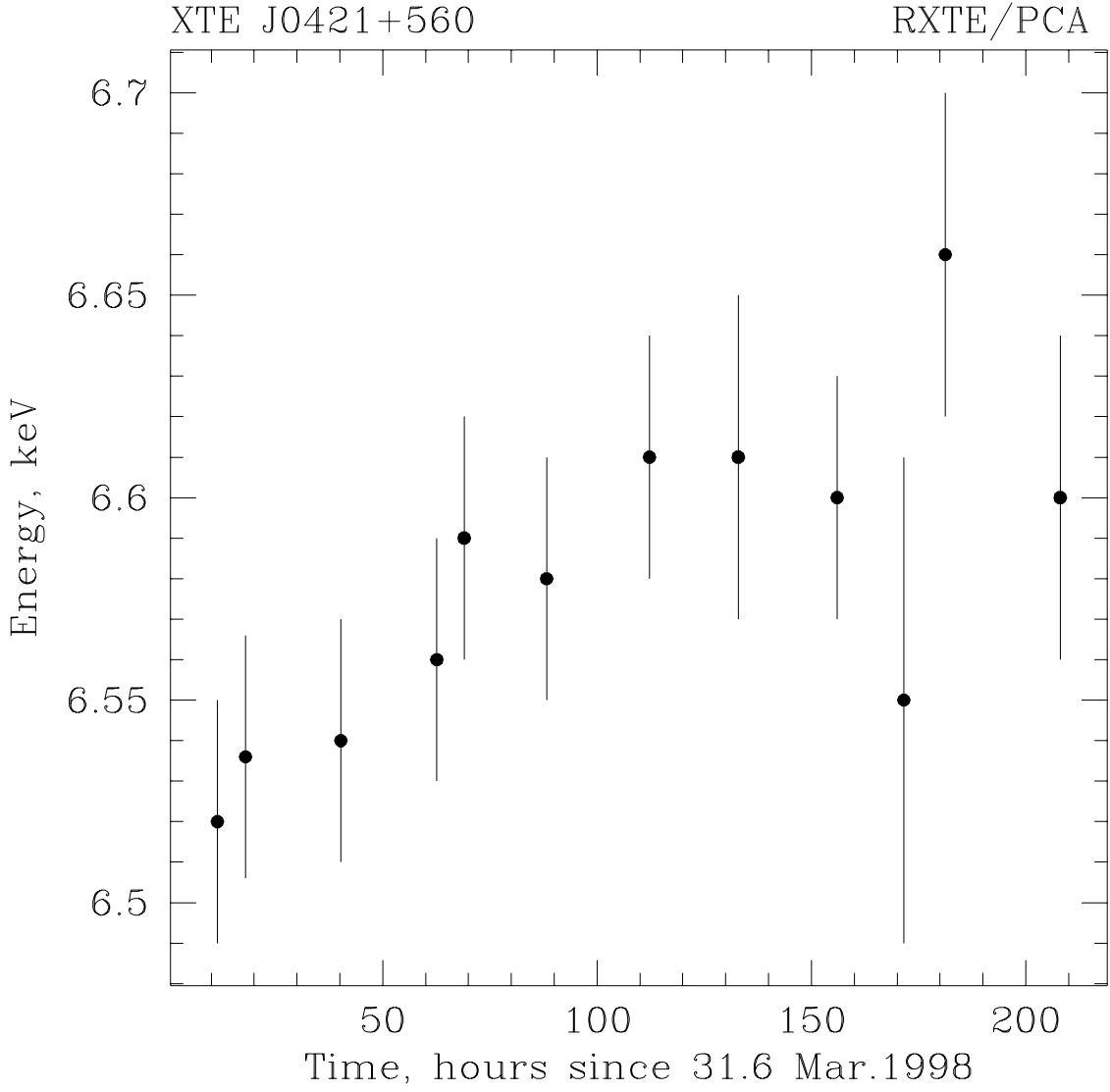}
\caption{Dependence of the Fe line central energy on time. \label{lines}}
\end{figure*}

\begin{figure*}
\epsfxsize=16cm
\epsffile{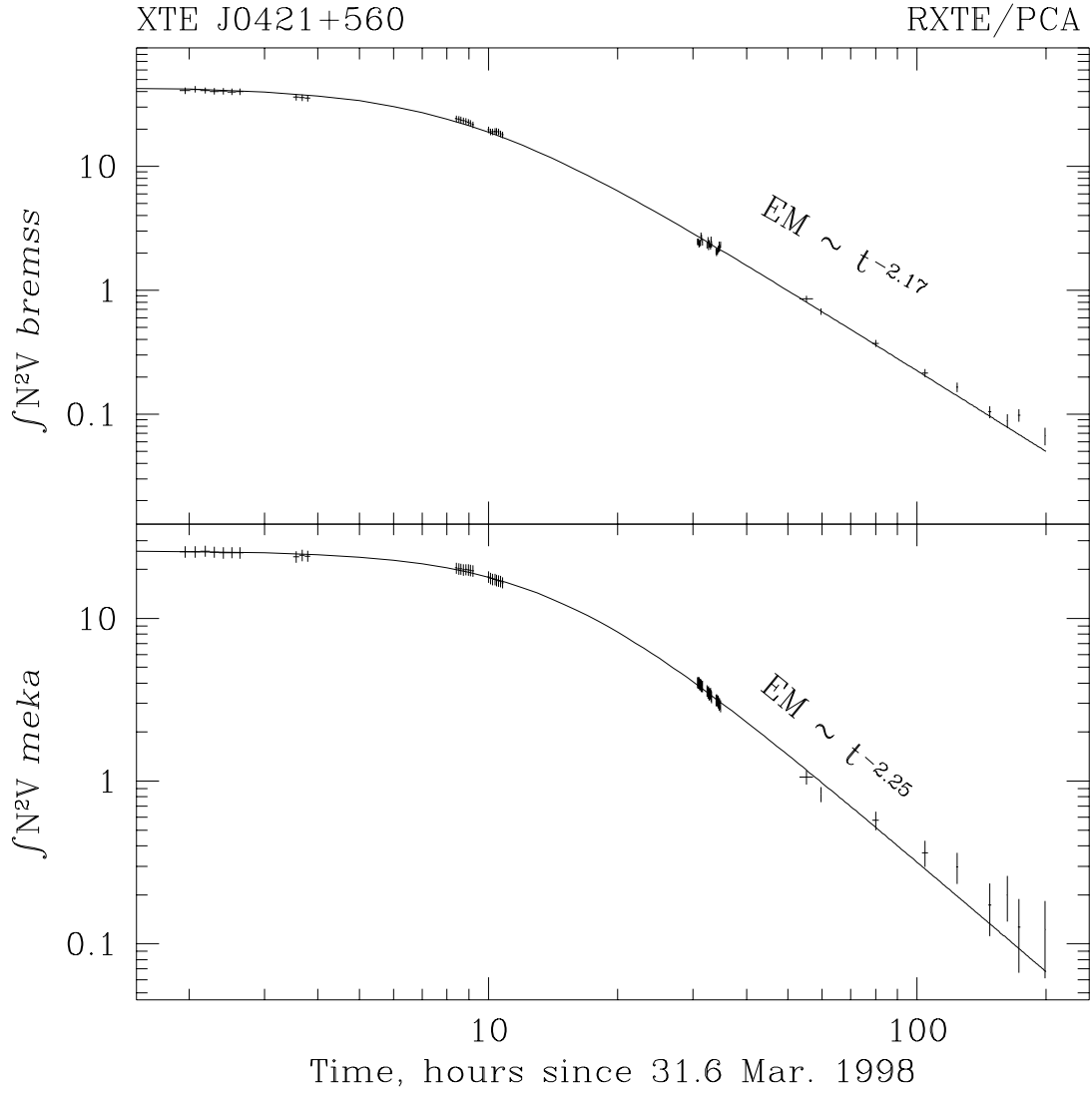}
\caption{Dependence of the Emission Measures on the time for spectral two
components. Solid lines show dependence $EM\sim
1/(1+(t/\tau)^{-\alpha})$ (see text). \label{dep_em}}
\end{figure*}

\begin{figure*}
\epsfxsize=16cm
\epsffile{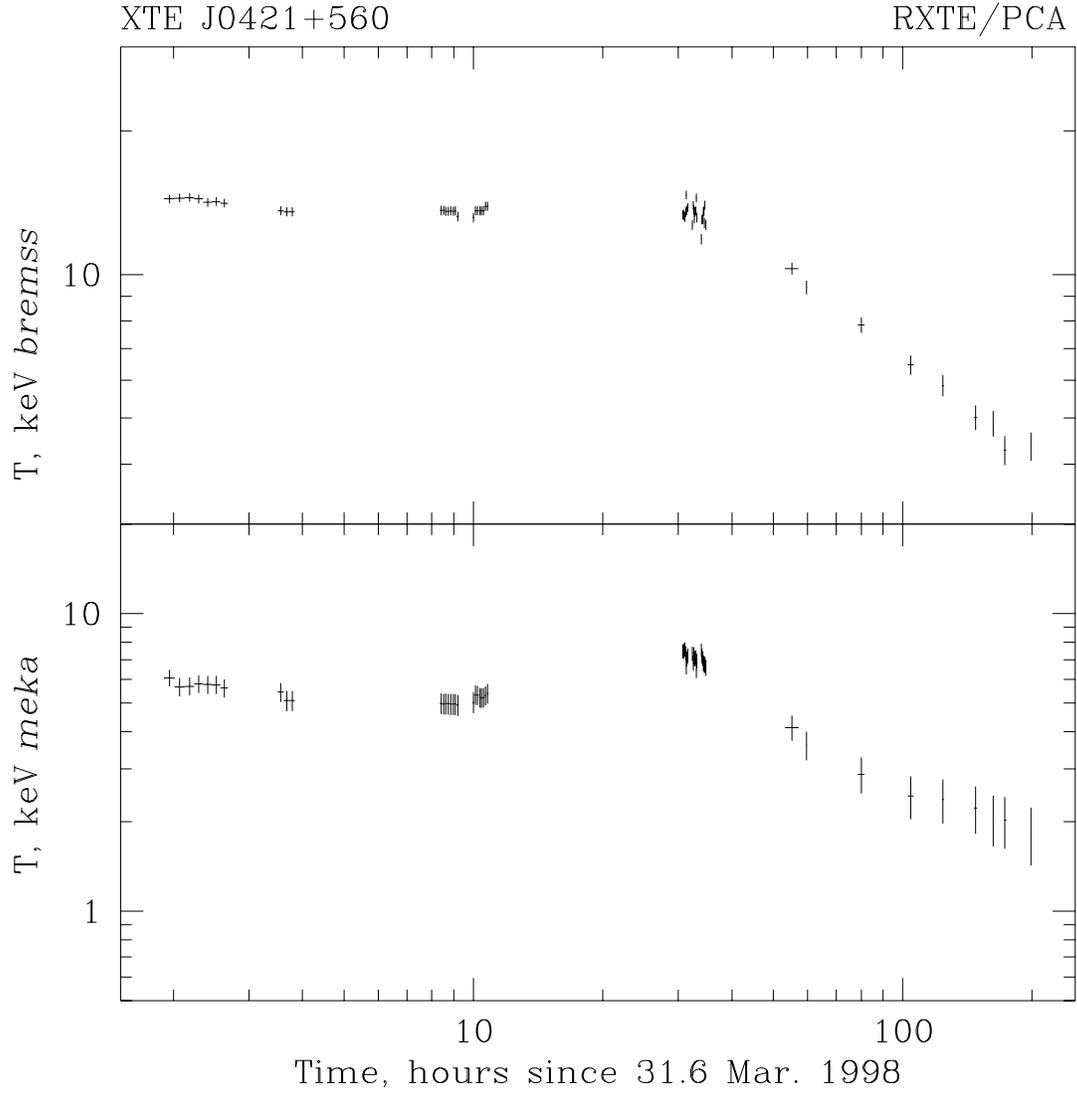}
\caption{Dependence of temperatures $T$ of two spectral components on time.\label{dep_t}}
\end{figure*}

\begin{figure*}
\epsfxsize=16cm
\epsffile{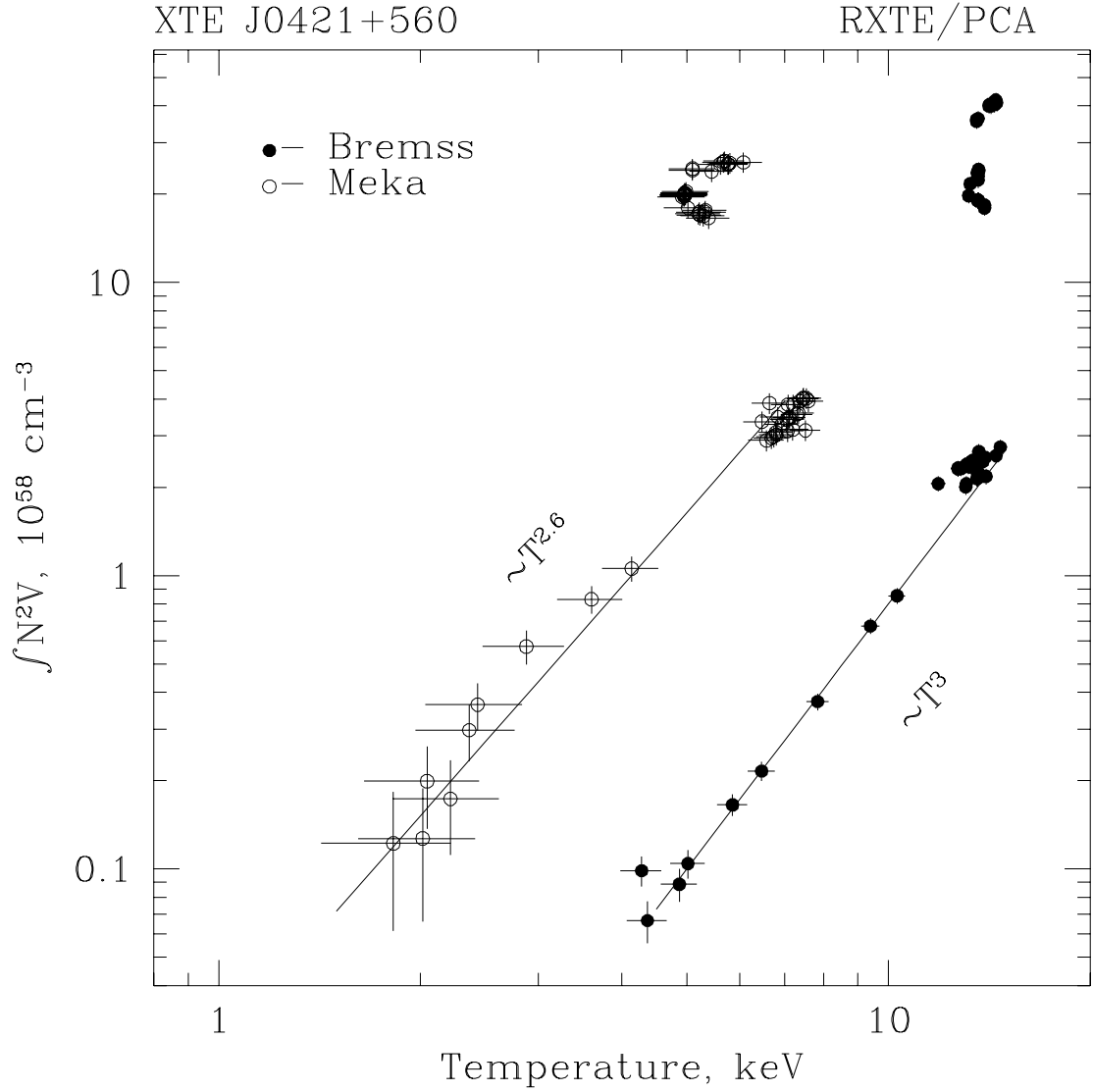}
\caption{Dependence of the Emission Measures on the temperatures of two
spectral components. \label{dep_em_t}}
\end{figure*}

\begin{figure*}
\epsfxsize=16cm
\epsffile{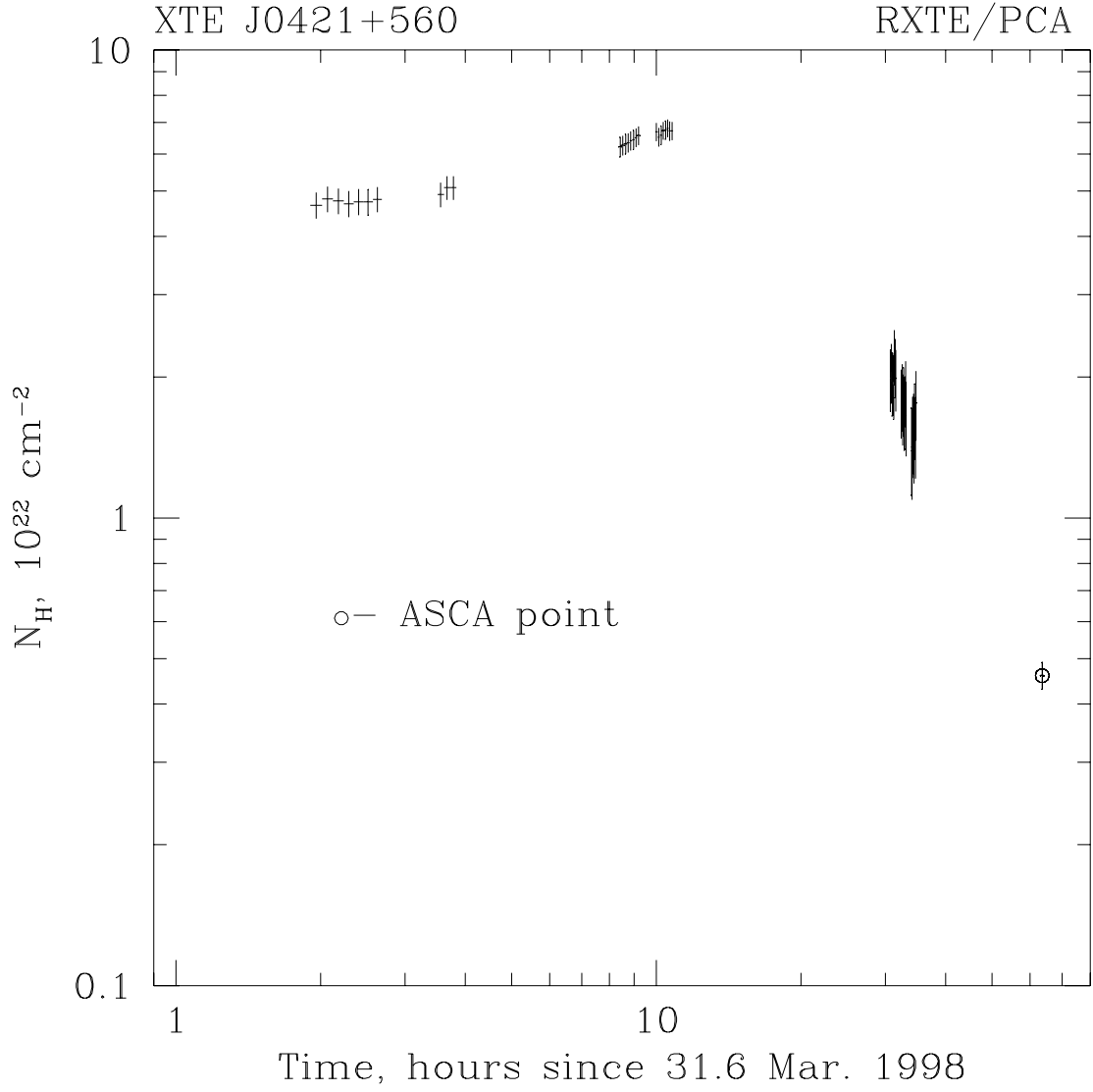}
\caption{Dependence of the low energy absorption value $N_H$ on time. \label{dep_nh}}
\end{figure*}

\clearpage

%TABLICY

\begin{center}

\begin{table}
\small
\caption{
RXTE observations of XTE J0421+560.\label{obslog}}
\begin{tabular}{clccccc}
\hline
\\
&Obs. ID& Date& Obs. time&\multicolumn{2}{c}{Exposure, sec.}\\
& & & &PCA&HEXTE$^a$\\
\hline
\\
1& 30171-01-01-00& 01/04/98& 01:52:48 - 03:52:32& 4109& 1443\\
2& 30171-01-02-07S& 01/04/98& 08:21:36 - 10:52:32& 6176& 4142\\
3& 30409-01-03-00& 02/04/98& 06:40:32 - 10:49:36& 9702& 3345\\
4& 30409-01-04-00& 03/04/98& 05:02:24 - 09:10:24& 2740& 892\\
5& 30409-01-04-01& 03/04/98& 11:30:24 - 11:50:24& 710& 217\\
6& 30409-01-05-00& 04/04/98& 06:41:20 - 09:27:28& 6950& 2283\\
7& 30409-01-06-00& 05/04/98& 06:44:48 - 10:10:24& 7590& 2671\\
8& 30409-01-07-00& 06/04/98& 03:21:36 - 04:30:56& 1503& 1016\\
9& 30409-01-08-00& 07/04/98& 02:25:20 - 05:01:36& 4181& 1389\\
10& 30409-01-08-01& 07/04/98& 18:02:08 - 18:22:56& 743& 191\\
11& 30409-01-09-00& 08/04/98& 03:35:12 - 06:11:28& 5830& 2009\\
12& 30409-01-10-00& 09/04/98& 06:33:20 - 07:32:32& 3035& 1059\\
\\
\hline
\end{tabular}
\begin{list}{}{}
\item[$^a$]- Dead time corrected exposure for each claster of HEXTE detectors.
\end{list}
\end{table}

\begin{table*}
\small
\caption{Parameters of approximation of XTE J0421+560 spectra by a power law
with high energy cutoff, two lines and absorption.\label{cutoff}}
\begin{tabular}{cccccc}
\hline
\\
&MJD&$\alpha$&$E_{cut}$&Flux (3--20 keV)& $N_H$\\
&&& keV&$10^{-10}$ erg/cm$^{2}$/s& $10^{22}$ atom/cm$^{2}$\\
\hline
\\
1& 50904.08&$ 1.53\pm0.05$&$ 12.1\pm0.5$&$ 578\pm11$&$ 4.8$\\
2& 50904.35&$ 1.59\pm0.04$&$ 11.9\pm0.4$&$ 350\pm7$&$ 6.5$\\
3& 50905.28&$ 1.33\pm0.04$&$ 9.0\pm0.3$&$ 59.1\pm1.2$&$ 1.5$\\
4& 50906.21&$ 1.69\pm0.04$&$ 9.3\pm0.3$&$ 16.2\pm0.3$&$ <1.0$\\
5& 50906.48&$ 1.78\pm0.05$&$ 9.2\pm0.6$&$ 11.3\pm0.2$&$ <1.0$\\
6& 50907.28&$ 1.98\pm0.05$&$ 8.6\pm0.4$&$ 5.6\pm0.1$&$ <1.0$\\
7& 50908.28&$ 2.2\pm0.1$&$ 8.0\pm0.5$&$ 2.74\pm0.05$&$ <1.0$\\
8& 50909.14&$ 2.0\pm0.1$&$ 6.1\pm0.5$&$ 2.01\pm0.08$&$ <1.0$\\
9& 50910.10&$ 2.1\pm0.4$&$ 5.4\pm0.6$&$ 1.07\pm0.07$&$ <1.0$\\
10& 50910.75&$ 2.4\pm0.4$&$ 7.2^{+1.4}_{-1.6}$&$ 0.85\pm0.06$&$ <1.0$\\
11& 50911.15&$ 2.0\pm0.4$&$ 4.6^{+0.7}_{-0.3}$&$ 0.75\pm0.06$&$ <1.5$\\
12& 50912.27&$ 2.2^{+0.3}_{-0.2}$&$ 5.1^{+0.9}_{-0.7}$&$ 0.55\pm0.06$&$ <1.0$\\
\hline
\\
\end{tabular}
\begin{list}{}{}
\item-- Parameter $N_H$ is presented without errors, because of the main
contribution to the uncertainty of this parameter is not the statistical
one, but rather the uncertainty of respose matrix on energy range
less than 4-5 keV. As usual the uncertainty of $N_H$ value is about $1\times
10^{22}$ 
\end{list}
\end{table*}

\begin{table*}
\small

\caption{The best fit parameters of the lines (PCA data).\label{lines}}
\begin{tabular}{ccccccccc}
\hline
\\
& & Blend 6.7 keV& & & Blend 8 keV& \\
\\
& E, keV& $\sigma$, keV& EW, eV& E, keV& $\sigma$, keV& EW, eV\\
\\
\hline
\\
1&$ 6.52\pm0.03$&$ 0.35\pm0.05$&$ 559\pm25 $&$ 8.19\pm0.14$&$0.1^a$&$ 77\pm16$ \\
2&$ 6.54\pm0.03$&$ 0.35\pm0.04$&$ 618\pm23 $&$ 8.20\pm0.14$&$0.1^a$&$ 82\pm14$ \\
3&$ 6.54\pm0.03$&$ 0.32\pm0.04$&$ 755\pm20 $&$ 8.19\pm0.15$&$0.1^a$&$ 69\pm18$ \\
4&$ 6.56\pm0.03$&$ 0.28\pm0.03$&$ 720\pm24 $&$8.07\pm0.15$&$0.1^a$&$ 93\pm17$ \\
5&$ 6.59\pm0.03$&$ 0.27\pm0.04$&$ 736\pm32 $&$ 8.32\pm0.15$&$0.1^a$&$ 102\pm26$ \\
6&$ 6.58\pm0.03$&$ 0.24\pm0.04$&$ 720\pm38 $&$ 8.04\pm0.15$&$0.1^a$&$ 110\pm27$ \\
7&$ 6.61\pm0.03$&$ 0.16\pm0.04$&$ 773\pm45 $& 8.10$^b$&$0.1^a$&$ <170$ \\
8&$ 6.61\pm0.04$&$ 0.23\pm0.04$&$ 794\pm61 $& 8.10$^b$&$0.1^a$&$ 163\pm80$ \\
9&$ 6.60\pm0.03$&$ 0.23\pm0.04$&$ 790\pm60 $& 8.10$^b$&$0.1^a$&$ 138\pm83$ \\
10&$ 6.55\pm0.06$&$ 0.35\pm0.1$&$ 1009\pm150$& 8.10$^b$&$0.1^a$&$ < 250$ \\
11&$ 6.66\pm0.04$&$ 0.21\pm0.1$&$ 717\pm61 $& 8.10$^b$&$0.1^a$&$ 174\pm78$ \\
12&$ 6.60\pm0.04$&$ < 0.25$&$ 674\pm61 $& 8.10$^b$&$0.1^a$&$ 177\pm88$ \\
\\
\hline
\\
\end{tabular}
\begin{list}{}{}
\item[$^a$]{Parameter $\sigma$ of $\sim$ 8 keV line was fixed on the value of
0.1, because of its unresolvable width.}
\item[$^b$]{Line position in sessions from 7 to 12 was fixed because of its weakness.}

\end{list}

\end{table*}

\begin{table*}
\small

\caption{Parameters of approximatin of PCA and HEXTE spectra by two component
model, consists of model of optically thin plasma without emission
lines ($bremsstrahlung$) and with emission lines ($meka$). \label{bremss_meka}}
\begin{tabular}{ccccccccc}
\hline
\\
\tiny
& $kT_{bremss}$& $\int{N^2dV}$& $kT_{meka}$& Red shift$^{a}$& $\int{N^2dV}$& $\chi^{2}$\\
& keV&            bremsshtr.&    keV&     &    meka&                     (290 dof.)\\
\hline
\\
1&$15.2\pm0.2$&$32.3\pm0.3$&$5.4\pm0.2$&$0.028\pm0.007$&$31.0\pm0.4$&324.68\\
2&$13.8\pm0.3$&$17.2\pm0.9$&$5.1\pm0.2$&$0.028\pm0.007$&$17.6\pm0.3$&298.46 \\
3&$13.5\pm0.3$&$2.48\pm0.08$&$5.9\pm0.2$&$0.029\pm0.007$&$3.67\pm0.04$&280.74 \\
4&$10.1\pm0.2$&$0.99\pm0.04$&$4.03\pm0.04$&$0.021\pm0.007$&$1.37\pm0.02$&241.21 \\
5&$9.4\pm0.3$&$0.65\pm0.02$&$3.6\pm0.1$&$0.017\pm0.007$&$0.89\pm0.03$&258.72 \\
6&$7.84\pm0.16$&$0.36\pm0.01$&$2.88^{+0.04}_{-0.09}$&$0.012\pm0.007$&$0.61\pm0.02$&269.72 \\
\\
\hline
\\
\end{tabular}
\small
\begin{list}{}{}
\item-- Emission measure is presented in units of $10^{58}$ (d/1kps)$^2$ cm$^{-3}$.
\item-- Absorbtion exists only in first, second and third sessions and equals
$\sim 4.8$, $\sim 6.6$  and $\sim 1.5$ in $10^{22}$ atom/cm$^{2}$ units respectivly.
\item[$^a$]-- Parameter error was fixed on 0.007 value assuming the absolute
precision of energy scale 0.05 keV. 
\end{list}

\end{table*}

\end{center}

\end{document}